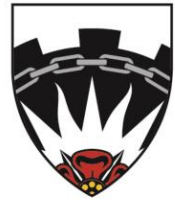

# Tailoring product ownership in large-scale agile

Bass, J and Haxby, A

# Tailoring Product Ownership in Large-Scale Agile

**Julian M. Bass**
*University of Salford, Manchester, UK*

**Andy Haxby**
*Comp001 IT BV, The Hague, Netherlands*

***ABSTRACT.** In large-scale agile projects, product owners undertake a range of challenging and varied activities beyond those conventionally associated with that role. Using in-depth research interviews from 93 practitioners working in cross-border teams, from 21 organisations, our rich empirical data offers a unique international perspective into product owner activities.*

*We found that the leaders of large-scale agile projects create product owner teams. Product owner team members undertake sponsor, intermediary and release plan master activities to manage scale. They undertake communicator and traveller activities to manage distance and technical architect, governor and risk assessor activities to manage governance. Based on our findings, we describe product owner behaviors that are valued by experienced product owners and their line managers.*

In large-scale Agile, product owners support multiple self-organizing teams that cooperate to build a shared product. As soon as self-organizing teams work together, they must sacrifice some level of autonomy. Feature delivery needs to be coordinated with other teams and often a project is a part of a portfolio of related development projects. Product owners have to cope with a range of new responsibilities, a wide range of stakeholders with conflicting needs and expanding workloads. In this context, our research has identified product owner role tailoring in which the role is no longer performed by a single individual but by a product owner team.

Conventionally, product owners are responsible for eliciting and prioritizing requirements and approving software produced for release to customers[1], see the activities in the center of Figure 1. We know that product owners must "spend time reshaping the product backlog"[2]. The product owner is the key Agile team member responsible for translating business needs into practical software requirements.

Practitioners in our study tell us that they evolve their own scaled Agile approaches by drawing on techniques from elsewhere, such as Disciplined Agile Delivery (DAD)[3] or Large Scale Scrum (LeSS)[4]. They are likely to have started out with the scrum-of-scrums approach, which is then adapted and tailored to fit their particular corporate quality standards and long-standing development process conventions. Every sprint retrospective creates opportunities to enhance and refine the agile processes used in teams.

We have identified three additional groups of activities that product owners in large-scale projects must perform around managing: scale, distance and governance, see Figure 1. Managing scale is concerned with handling a large number of stakeholders and long term software release timescales; managing distance is concerned with global software development and managing governance relates to achieving appropriate consistency among cooperating agile teams.

Our empirical research is based on research interviews with 93 practitioners from 21 companies over an 8 year period, see research methods sidebar for more details. We consider large-scale to consist of at least 25 developers configured into more than 3 cooperating teams working together for a period of more than 9 months. All the teams in our study had to contend with geographical distribution, which we refer to as cross-border teams. We will discuss each of the three sets of product owner activities and recommend behaviors that are valued by experienced product owners and their line managers.

## Managing Scale

As mentioned earlier, product owners, in large-scale Agile, have to serve larger groups of cooperating teams over longer timescales. Consequently, they manage relationships with a wider range of stakeholders. To achieve this, product owners in our study perform more networking and expectation setting than a single individual can cope with.

### Product Sponsor

The product sponsor develops the vision as well as creating and negotiating a business case. Either selling a large product, winning a large tender or approving an enterprise-scale internal project usually requires most senior board level involvement. While a CEO, CIO or CTO may "own" the project, they are unlikely to have time to attend to project detail. They have a large organization to run, after all. As one very senior civil servant in our study said "my biggest challenge, in building digital systems, is translating [Government] policy into real outcomes." Product sponsors surround themselves with a product owner team and delegate to a named product owner (see intermediary, below). However, we found that product sponsors maintain focus on the project vision by reviewing important demonstrations, see the behaviors in Table 1.

### Intermediary

An intermediary interfaces with the product sponsors and coordinates or negotiates with key project stakeholders. Intermediaries have a sound understanding of the goals of the project obtained by regular meetings with the product sponsor. Nevertheless, the purpose of the intermediary is to be more accessible and available to other project stakeholders than product sponsors. However, the intermediaries in our study invite product sponsors to key demonstrations and solicit their feedback.

### Release Plan Master

The release plan master manages a release pipeline to synchronize cooperating teams. Large-scale projects tend to operate over several years and must be coordinated around multiple milestones. One of the companies in our study manufactures various types of body scanner used in hospitals. Such medical instruments are highly regulated but also the machines have a long working life. The manufacturer must support the software used in such scanners for 20 years or more.

Product owners in our study plan major releases to coincide with external factors such as TV advertising schedules. Release dates are included in planning as a type of project feature. As advertising deadlines approach, other project features are de-prioritised to meet the delivery objectives. While specific feature details are not defined far in advance, when using Agile methods, the main goals are planned several releases ahead.

## Managing Distance

Product owners in large-scale agile routinely participate in geographically distributed software development projects[5]. They have to manage physical distance, temporal distance and cultural distance. In response, product owners form themselves into teams to support and engage stakeholders at different locations and sites.

### Communicator

The communicator uses multi-media technologies to connect onshore, offshore and others as a consequence of geographical distribution, where face-to-face communication is not possible. Product owners seem to prefer video and audio conferencing while scrum masters and developers in our study seem to prefer instant messaging. Experienced product owners select appropriate communication media for their intended audience, see the behaviors in Table 1.

### Traveler

The traveler actually spends time onshore or offshore. Face-to-face communication remains the gold standard for building trust, empathy and understanding. Members of the product owner team travel to remote sites to build trust, learn about local conventions and motivations, and inculcate a shared ethos. Travelers spend months with clients and development teams soaking up the atmosphere and understanding local pressures.

## Managing Governance

Where self-governing teams must cooperate, it becomes necessary to share a common set of standards for quality and technology[6]. Self-organizing teams must sacrifice some creativity and autonomy to reach consensus on common standards. The level of governance is also a project feature to be balanced against functional scope. Experienced product owners in our study manage to encourage creativity and innovation while also ensuring compliance with shared standards and guidelines.

### Technical Architect

The organizations in our study, co-opt technical specialists onto the product owner team to provide architectural coordination among co-operating teams. The mix of technologies used on larger scale projects can often become complex. Further, the larger number of teams can expect to have staff departing and joining during projects providing a risk to architectural coherence. It is often desirable to disseminate best practice across teams. Architects adopt architectural styles, develop reference architectures and select design and deployment patterns. These architectural structures need to be unobtrusively disseminated but consistently applied. This requires architects, supporting the product owner team, with technical stature and excellent influencing skills, see the behaviors in Table 1.

### Governor

The governor ensures that the project complies with corporate quality standards and technical policies. Self-organizing teams relinquish some autonomy towards an architecture board or design authority that determines common policies and approaches. Product owners manage relationships with such boards in order to understand constraints placed on teams but also to influence and perhaps initiate change as desirable new technologies become sufficiently established.

### Risk Assessor

The risk assessor, in large-scale projects, evaluates technical complexity, lists risks and plans mitigation. The product owners in our study argue that the large sums of money, risk of adverse publicity and reputational damage and negative consequences of project failure make it attractive to actively monitor and manage risk[6]. For example, product owners, at one global software service provider, establish and update a risk assessment log by conducting risk assessments for each team as part of every sprint. This precautionary approach allows early identification of emerging threats to project success and enables early mitigation.

## Product Owner Behaviors

We have summarized key behaviors for product owners, mapped to the various activities we found, in Table 1. However, four general themes arise from the teams in our study.

### Favor Face-to-Face Interactions

When dealing with geographical, temporal and cultural distance it is tempting to fall back on written communication using email and word documents. We found that product owners view these tools as superficially effective. Understanding, trust and empathy come from building social capital through face-to-face interactions. Product owners that fiercely network with clients, scrum masters and other stakeholders seem to have more successful project outcomes[8].

### Understand and Focus on Real Goals

On projects regarded by practitioners as successful, product owners appear to use influencing skills to keep a wide range of stakeholders targeted on a specific and focused set of goals. Objective test criteria enable teams to demonstrate progress towards project goals. Practitioners value product owners who stay true to key project goals even as impediments and obstacles arise. The ability to keep the "big picture" in mind, even when pivoting around challenges and resource constraints, is indeed a skill.

### Use a Minimum Viable Product to Permit Change

On complex projects, experienced product owners find a minimum viable product[9] (MVP) valuable for enabling change. The MVP is not the final system, so it can gain traction in overcoming resistance to change. The MVP can pilot new development processes, introduce new technologies and explore new approaches to governance. Approval of the MVP can then provide permission to embed these new ideas.

**Make Product Owner Teams Explicit**

The product sponsor, intermediary, technical architect and other product owner team members form a product owner team[10]. The product owner team members perform a wide range of activities, as shown in Figure 1. We argue it is helpful to make building the product owner team explicit. Sponsors should think about team building, induction of new members and succession planning.

**Conclusions**. The product owners in our study perform activities to manage scale, distance and governance as part of a product owner team. We argue for prioritising face-to-face interactions, maintaining focus on goals and the knowing creation and sustenance of a product owner team to support project stakeholders.

**Dr Julian M. Bass** is a Senior Lecturer (Associate Professor) in software engineering at the University of Salford, Manchester, UK. Contact him at j.bass@salford.ac.uk.

**Andy Haxby** Andy Haxby is Director of Dutch software services company Competa and founder of the Fair Trade Software Foundation. Contact him at andy@competa.com.

## Research Methods

This article draws on over eight years' of research conducted specifically investigating product ownership in large-scale cross-border software development. We have observed teams performing sprint planning, daily status (stand-up meetings), product demonstrations and retrospectives. We have also examined public and commercially confidential project and corporate documents from the companies in our study. Such documents have included development processes, governance policies, design documents and test plans.

The main source of research data has been from 93 practitioner interviews, from 21 companies and UK government organizations. The organizations in the study include well known multinational internet and software service companies as well as government agencies, and companies in the retail, CRM and banking/finance sectors.

The interviewers employed a semi-structured interview guide[12] which included open-ended questions to elicit topics from respondents not considered by the interviewer. Respondents include product owners and their line managers (often a CIO, CTO or head of engineering), as well as middle managers, Agile coaches and development team members such as software developers, testers and scrum masters.

Interviews have been recorded, transcribed and analyzed employing a Glaserian grounded theory[13] approach. Open coding, "memoing," constant comparison and theoretical sampling are used to extract topics, concepts and themes from the interview transcript data.

*Table 1 Behaviors for Managing Large-scale, Distance & Governance*

| Product Owner Roles and Activity Names | Typical Artefacts[a] | Good Behaviors | Behaviors to Avoid |
|---|---|---|---|
| Product sponsor | Core project goals and vision | Relentlessly focuses on key goals<br><br>Defines clear requirements<br><br>Makes time for reviewing product, e.g. attending important demos | Hands-off problems<br><br>Relies on documents<br><br>Delegates challenges and believes in magic |
| Intermediary | Project goals and vision | Understands and trusts agile<br><br>Connects all the right people | Interferes, wants to put own stamp on everything. |
| Release Plan Master | Release plans | Understands and minimizes dependencies<br><br>Pushes for appropriate workloads with realistic release plans | Fails to balance customer needs, scope and technical debt |
| Communicator | Slack, blog and wiki posts | Communicates effectively, uses appropriate channels e.g. slack and other instant messaging for developers; Trello for scrum masters | Communicates everything by email |
| Traveler | Slack, blog and wiki posts | Makes enough time for extensive networking<br><br>Favors face-to-face interaction. | Avoids face-to-face networking, relies on documents and email |
| Technical Architect | Reference architecture | Networks with scrum masters and architecture board influencers<br><br>Is approachable and communicative<br><br>Focuses on people AND technology | Stays in the background<br><br>Focuses on technology<br><br>Doesn't build relationships with key stakeholders |
| Governor | Quality standards | Trusts agile, inputs important requirements to PO team<br><br>Attends key demos | Focuses excessively on administrative aspects of quality assurance |
| Risk Assessor | Risk register | Inside team to understand project goals and status | Focuses excessively on administrative aspects of risk management |

[a] Adapted from [7]

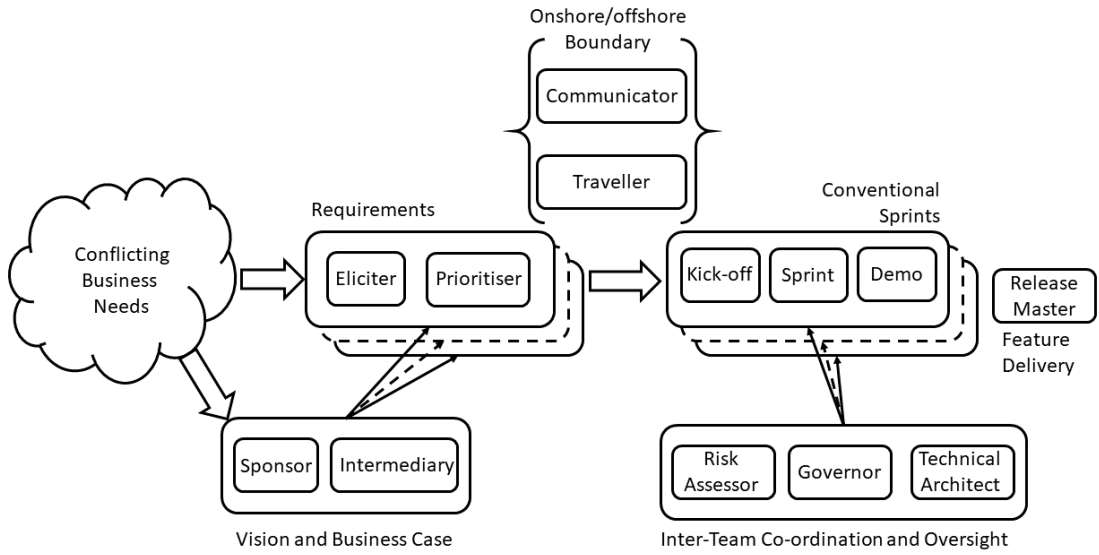

Figure 1 Product Owner Activities (Adapted from[10])